\documentclass{aa}
\usepackage{graphics}

\newcommand{\p}[2]{\frac{\partial#1}{\partial#2}}

\begin{document}
\thesaurus{12(02.08.1; 02.19.1; 09.10.1; 09.13.2)}
\title{On the transfer of momentum from stellar jets to molecular outflows}
\author{T.P. Downes \inst{1, 2}
\and T.P. Ray \inst{3}}
\offprints{T. P. Downes}
\mail{School of Mathematics, Trinity College, Dublin 2, Ireland}
\institute{Sterrenkundig Instituut, Postbus 80000, 3508 TA Utrecht, The
Netherlands
\and School of Mathematics, Trinity College, Dublin 2, Ireland
\and Dublin Institute for Advanced Studies, 5 Merrion Square, Dublin 2,
Ireland}
\date{Received date ;accepted date}
\maketitle

\begin{abstract}

While it is generally thought that molecular outflows from 
young stellar objects (YSOs) are accelerated by underlying stellar winds or 
highly collimated jets, the actual mechanism of acceleration remains uncertain. 
The most favoured model, at least for low and intermediate mass stars, is that 
the molecules are accelerated at jet-driven bow shocks. Here we investigate, 
through high resolution numerical simulations, the efficiency of this mechanism
in accelerating ambient molecular gas {\em without causing dissociation}. The 
efficiency of the mechanism is found to be surprisingly low suggesting that 
more momentum may be present in the underlying jet than previously 
thought. We also compare the 
momentum transferring efficiencies of pulsed versus steady jets. We find 
that pulsed jets, and the corresponding steady jet with the same average 
velocity, transfer virtually the same momentum to the 
ambient gas. The additional momentum ejected sideways from the jet beam 
in the case of the pulsed jet only serves to accelerate post-shock {\em jet} 
gas which forms a, largely atomic, sheath around the jet beam.  

For both the steady and pulsing jets, we find a power law relationship between 
mass and velocity ($m(v) \propto v^{-\gamma}$) which is similar to what is 
observed. We also find that increasing the molecular fraction in the jet 
decreases $\gamma$ as one might expect. We reproduce the so-called Hubble law 
for molecular outflows and show that it is almost certainly a local effect 
in the presence of a bow shock.

Finally, we present a simple way of overcoming the numerical problem of
negative pressures while still maintaining overall conservation of
energy.

\end{abstract}

\section{Introduction}
\label{introduction}
     It has been proposed that molecular outflows, at least from low and 
intermediate mass young stars, may be driven by highly collimated 
jets (see, for example, Padman, Bence \& Richer \cite{PBR}). Although the 
likely mechanism by which such jets transfer their momentum to the ambient 
medium remains unknown, a number of ideas have been put forward 
(for a review of models the reader is referred to Cabrit, Raga \& Gueth 
\cite{CRG}). Of these the most promising seems to be the so-called ``prompt
entrainment'' mechanism. According to this model the bulk of the molecular 
outflow is accelerated ambient gas near the head of the jet or more 
precisely along the wings of its associated bow shock. Observational support 
for prompt entrainment comes from the spatial coincidence of shocked molecular 
hydrogen bows with peaks in the CO outflow emission (e.g.\ Davis \& 
Eisl\"offel \cite{D&E}). 

Smith, Suttner \& Yorke (\cite{smith}) and Suttner et al.\ (\cite{Suttner}) 
have carried out a number of 3-D simulations of dense molecular jets 
propagating into a dense medium in order to test the prompt entrainment  
hypothesis. These authors found that their simulations reproduced many of the 
observational characteristics of molecular flows including the so-called 
`Hubble law'  (see, e.g. Padman et al.\ \cite{PBR}) and strong forward, as 
opposed to sideways, motion (Lada \& Fich \cite{LF}). While such results are 
encouraging for jet-driven models, it is still fair to say that no individual 
model has yet been able to plausibly account for all the observations (Lada \& 
Fich \cite{LF}).  Moreover, alternatives to the jet model may be better 
at explaining the observational characteristics of some molecular flows 
(e.g.\ Padman et al.\ \cite{PBR} and Cabrit et al.\ \cite{CRG}).

The limited resolution of the 3-D jet simulations of Smith et al.\ 
(\cite{smith}), and Suttner et al.\ (\cite{Suttner}), along with the high 
densities used by these authors, meant that they could not resolve the
post-shock cooling regions in the flow. In addition it was not possible to 
explore parameter space as only a few such simulations could be 
performed. Here we take a somewhat different approach by assuming low 
density atomic/molecular jet mixtures, a low density ambient medium and 
cylindrical symmetry. Although such an approach obviously has it 
limitations, it does allow us to explore parameter space more fully and 
to resolve post-shock cooling regions (this might be important, for 
example, if one is to gauge the importance of certain instabilities). The 
primary goal of this work is to investigate the efficiency of YSO jets in 
accelerating ambient molecular gas {\em without causing dissociation} of 
its molecules.  

An additional question we address in this paper is whether velocity 
variations (pulsing) of the jet might enhance transfer of momentum from the 
jet to its surroundings and thus help to accelerate ambient gas. Pulsing
induces internal shocks which can squeeze jet gas sideways (Raga et al.\ 
\cite{Retal93}). This gas does not interact with the 
ambient medium directly, but is instead squirted into the cocoon of processed 
(post-shock) jet gas, which separates the jet from the ``shroud'' 
of post-shock ambient gas. Chernin \& Masson (\cite{C&M}) however argue that, 
through the cocoon, momentum from the jet may be continuously coupled to  
the ambient flow. 

The properties of the simulated systems in which we are interested are as
follows:
\begin{itemize}
\item How much momentum is transferred to the ambient molecules?
\item Is there a power-law relationship predicted between mass in the 
molecular flow and velocity?
\item What are the proper motions of the molecular `knots', and how does
their emission behave with time?
\item Is the so-called `Hubble law' of molecular outflows reproduced under 
reasonable conditions?
\item Is there extra entrainment of ambient gas along the jet due to 
velocity variations?
\end{itemize}

We will discuss each of these points in turn when presenting our results. 

Our numerical model is presented in \S 2 and our results in \S 3. Conclusions 
from this work are presented in \S 4 and a simple way of overcoming the 
numerical problem of negative pressures while still maintaining overall
energy conservation is given in the Appendix.

\section{Numerical model}

\subsection{Equations and numerical method}
\label{num_method}

The equations solved are
\begin{eqnarray}
\p{\rho}{t} & = & - \vec{\nabla} \cdot (\rho \vec{u}) \label{contin}\\
\p{\left(\rho \vec{u} \right)}{t} & = & - \vec{\nabla}\cdot[ \rho
\vec{u}\vec{u}+ P\vec{\vec{I}} ] \label{mom}\\
\p{e}{t} & = & - \vec{\nabla}\cdot\left[\left(e+P\right)\vec{u}\right]-L
\label{energy} \\
\p{n_{\rm H} x}{t} & = & -\vec{\nabla}\cdot\left[n_{\rm H} x
\vec{u}\right]+ J(x,n_{\rm H},T)
\label{ifrac} \\
\p{n_{\rm H_2}}{t} & = & - \vec{\nabla} \cdot (n_{\rm H_2} \vec{u}) -
n_{\rm H_2}n_{\rm H}k(T) \label{h2} \\
\p{n_{\rm H}}{t} & = & - \vec{\nabla} \cdot (n_{\rm H} \vec{u}) +
2n_{\rm H_2} n_{\rm H} k(T) \label{h} \\
\p{\rho \tau}{t} & = & - \vec{\nabla} \cdot (\rho \tau \vec{u})
\label{tracer}
\end{eqnarray}
\noindent where $\rho$, $\vec{u}$, $P$, $e$ and $I$ are the mass density,
velocity, pressure, total energy density and identity matrix respectively.
$n_{\rm H}$ and $n_{\rm H_2}$ are the number densities of atomic and molecular 
hydrogen, $x$ is the ionization fraction of atomic hydrogen, $T$ is the 
temperature, $J(x,n_{\rm H},T)$ is the ionization/recombination rate of 
atomic hydrogen, $k(T)$ is the dissociation coefficient of molecular 
hydrogen, and $\tau$ is a passive scalar which is used to track the jet 
gas.  We also have the definitions
\begin{eqnarray}
e&=&\frac{1}{2}\rho \vec{u}\cdot\vec{u} + \frac{c_v}{k_{\rm B}} P \\
L&=&L_{\rm rad}+E_{\rm I} J(x,n_h,T)+E_{\rm D} k(T) 
\end{eqnarray}
\noindent where $c_v$ is the specific heat at constant volume, 
$k_{\rm B}$ is Boltzmann's constant, $E_{\rm I}$ is the ionization
energy of hydrogen and $E_{\rm D}$ is the dissociation energy of H$_2$.  
So $L$ is a function which denotes the energy loss and gain due to 
radiative and chemical processes.  $L_{\rm rad}$ is the loss due to 
radiative transitions and is made up of a function for losses due to 
atomic transitions (Sutherland \& Dopita \cite{S&D}), and one for losses 
due to molecular transitions (Lepp \& Shull \cite{L&S}).  The second term
in $L$ is the energy dumped into ionization of H, and the third is that 
dumped into dissociation of H$_2$.  The dissociation coefficient 
$k(T)$ is obtained from Dove \& Mandy (\cite{D&M}) and the ionization 
rate, $J$, is that used by Falle \& Raga (\cite{F&R}).

These equations are solved in a 2D cylindrically symmetric geometry
using a temporally and spatially second order accurate MUSCL scheme (van
Leer \cite{leer}; Falle \cite{falle}).  The code uses a linear 
Riemann solver except where the resolved pressure differs from either the 
left or right state at the cell interface by greater than 10\% where it uses 
a non-linear solver (following Falle 1996, private communication).  Non-linear 
Riemann solvers allow correct treatment of shocks and rarefactions without 
artificial viscosity or entropy fixes.  Applying them only in non-smooth 
regions of the flow means that, while the benefits are the same, the 
computational overhead is minimised.  This code is an updated version of 
that described in Downes \& Ray (\cite{D&R}).

Sometimes negative pressures are predicted by simulations involving
radiative cooling.  Typically these are overcome by simply resetting the
calculated pressure to an arbitrary, but small, positive value.  However,
this involves injecting internal energy into the system and this is 
undesirable.  A fairly reliable way of overcoming this problem is discussed 
in Appendix A.

\subsection{Initial conditions}

Initially the ambient density and pressure on the grid are uniform and
defined so that the ambient temperature on the grid is $10^2$ K.  The jet
temperature is set to $10^3$ K.  The function $L$ is set to zero below 
this latter temperature as the data used in the cooling functions becomes 
unreliable and cooling below this temperature is not dynamically
significant anyway.  In most cases the ratio of jet density to ambient density
($\equiv \eta$) is set to 1 (see Table \ref{list_sims}).  The ratio
$\frac{n_{\rm H_2}}{n_{\rm H}}=9$ both inside and outside the jet,
unless otherwise indicated (again, see Table \ref{list_sims}).  In all
cases the gas is assumed to be one of solar abundances.  The boundary 
conditions are reflecting on $r=0$ (i.e.\ the jet axis) and on $z=0$ except 
where the jet enters, and gradient zero on every other boundary.  
The computational domain measures $1500\times300$ cells (but larger in 
the $\eta=10$ simulations), with a spacing of $1\times10^{14}$ cm.  We 
find that 
the efficiency of momentum transfer is sensitive to the grid spacing.  
We performed a number of simulations with different spacings and 
concluded that this is the absolute minimum necessary to get reliable 
results.  This length should be reduced with increasing density.  Incidentally
this means that examining this property at the densities used by, 
for example, Smith et al.\ (\cite{smith}) is impractical.  

The jet enters the grid at $z=0$ and $r\leq R$ and the boundary conditions 
are set to force inflow with the jet parameters.  $R$ is set at
$5\times10^{15}$ cm or 50 grid cells.  The jet velocity is given by
\begin{equation}
\label{jet_noshear}
v_{\rm jet}(t)=v_0+\frac{v_1}{4}\sum_{j=1}^4 \sin(\omega_j t)
\end{equation}
with $v_0\approx215$ km s$^{-1}$ corresponding to a Mach number of 65
and $\omega_j$ are chosen so that the corresponding periods are 5, 10, 20 
and 50 yrs.  Here $v_1$ is effectively an amplitude for the velocity 
variations where present. The jet is initially given a small shear layer of 
about 5 cells ($5\times10^{14}$ cm) in order to avoid numerical problems at the
boundary between the jet and ambient medium.  In this layer the velocity
decays linearly to zero.

Nine simulations were run with varying values of density and
velocity perturbation.  These are listed in Table \ref{list_sims} which 
also gives the key we will use to refer to the simulations.  In addition
simulations of a purely atomic jet, and of a jet with a wide shear layer of 
almost 25 cells ($2.5\times10^{15}$ cm) were run.  The velocity of the 
jet with the wide shear layer is given by
\begin{equation}
v_{\rm jet}(t, r)=\frac{v_{\rm jet}(t)}{2}\left\{1-\tanh\left(\frac{r-R}
{10^{15} \rm{ cm}} \right)\right\}
\end{equation}
where $v_{\rm jet}(t)$ is given by Eq. \ref{jet_noshear}. 
Each system was simulated to an age of 300 yrs.  Although this 
is very young compared to the observed age of stellar jets, it was felt 
that the qualitative behaviour of the system at longer times could reliably
be inferred from these results.  The densities chosen are rather low to 
ensure adequate resolution of the system as described above.  Unfortunately 
this precludes the use of the data of McKee et al. (\cite{mckee}) for 
calculations of the emissions from CO as their calculations are only 
valid for gases of much higher densities.  As a result we present our 
findings in terms of the mass of molecular gas rather than its 
luminosity.

\begin{table}[t]
\begin{tabular}{cccc}\hline
Label & Jet density (cm$^{-3}$)& ($\frac{v_1}{v_0}$) & Notes \\ \hline
A & 10 & 0 & - \\
B & 10 & 0.6 & - \\
C & 100 & 0 & - \\
D & 100 & 0.1 & - \\
E & 100 & 0.2 & - \\
F & 100 & 0.4 & - \\
G & 100 & 0.6 & - \\ 
G1 & 100 & 0.6 & Wide shear layer \\
G2 & 100 & 0.6 & Atomic jet \\ 
H & 100 & 0.0 & $\eta=10$ \\
I & 100 & 0.6 & $\eta=10$ \\ \hline
\end{tabular}
\caption{\label{list_sims} A list of the simulations performed in this
work.  Unless otherwise stated $\eta=1$}
\end{table}

\section{Results}

\begin{figure*}
\resizebox{\hsize}{!}{\includegraphics{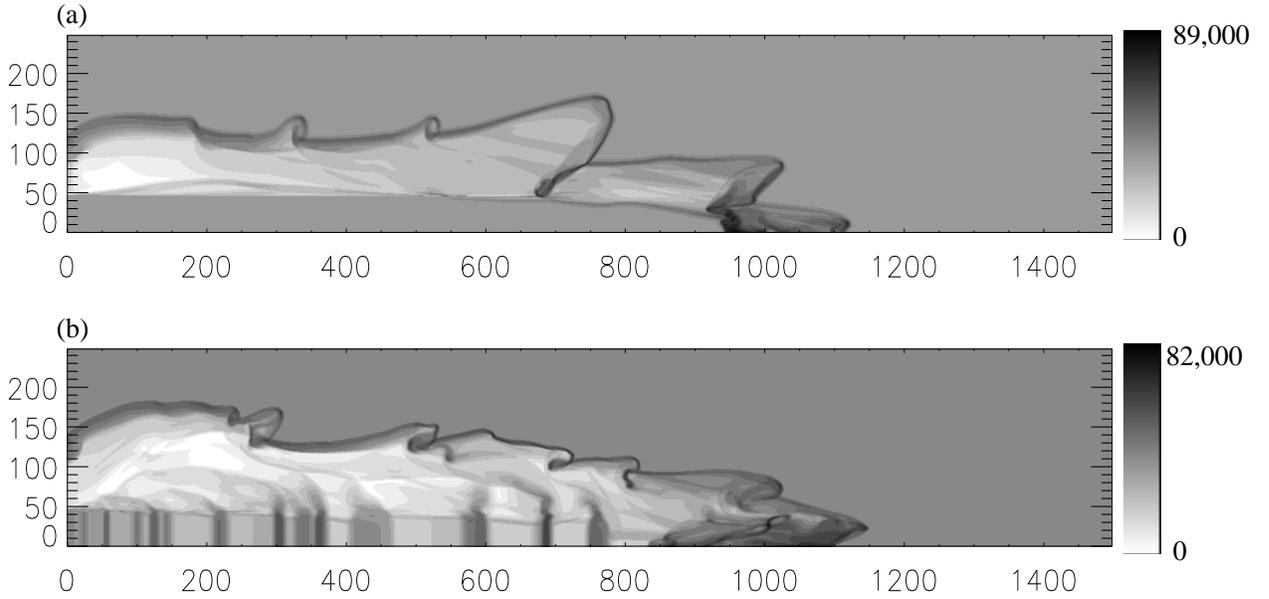}}
\caption{Log-scale plots of the distribution of number density for (a), 
simulation C, and (b), simulation G at $t=300$ yrs.  The scales are in 
units of cm$^{-3}$}
\label{densities}
\end{figure*}

Fig.\ \ref{densities} shows plots of the distribution of number
density for simulations C and G.  The cocoon of the varying jet has many 
bow-shaped shocks travelling through it as a result of the internal working 
surfaces in the jet forcing gas and momentum out of the jet beam.  It is
interesting to note that the bow shock of the steady jet is more irregular 
than that of the pulsed jet.  This irregularity is probably due to the 
Vishniac instability (e.g.\ Dgani et al.\ \cite{dgani}) growing at the 
head of the jet.  Presumably the variation of the conditions at the head 
of the varying jet dampens the growth of this instability.  It should be
noted that the enforced axial symmetry in these calculations makes the
bow shock appear more smooth than it would in 3D simulations.  However,
the phenomenon of the bow shock breaking up occurs in both 2D and
3D simulations.

We will discuss each of the properties mentioned in \S
\ref{introduction} below.

\subsection{Momentum transfer}

We make use of the jet tracer $\tau$ to track how much
momentum has been transferred from the jet to the ambient medium.  The
fraction of momentum transferred from jet gas to ambient molecules is
\begin{equation}
{\cal F}_{\rm H_2}=\frac{\sum_{i,j}(1-\tau_{ij})|\vec{u}_{ij}|
m_{\rm H_2}n_{{\rm H_2},ij} dV_{ij}}{\sum_{i,j} |\vec{u}_{ij}|n_{ij}<m>dV_{ij}}
\end{equation}
\noindent where $n_{{\rm H_2},ij}$ and $n_{ij}$ are the number density
of molecular hydrogen and the total number density in cell $ij$
respectively, $i$ and $j$ are cell indices in the $z$ and $R$
directions, and $dV_{ij}$ is the volume of cell $ij$.  This equation is
valid since the only momentum on the grid originated in the jet, and
since the simulations are stopped before any gas flows off the
grid.  Note that we only consider momentum transferred to ambient
molecules because we are only interested in how efficient YSO jets
are at accelerating molecules, not atoms.  Thus we ignore ambient
molecules which have been dissociated in the acceleration process.
Table \ref{fractions} shows the fraction of momentum in ambient
molecules for all the simulations after 300 yrs.  For completeness we 
also show ${\cal F}_{\rm total}$, the total fraction of momentum 
transferred to ambient gas, whether molecular or atomic.

The first interesting point to note from Table \ref{fractions} is that 
the amount of momentum residing in ambient molecules in these 
simulations is typically an order of magnitude less than the total 
momentum contained on the grid. Note, however, how significant amounts 
of momentum are transferred to the ambient medium {\em as a whole}, 
especially in those cases where the jet density matches that of its 
environment. This is as one would expect. What is perhaps surprising at first
is the low efficiency of momentum transfer to ambient gas that remains
in molecular form in the post-bow shock zone. 
 
Comparison between ${\cal F}_{\rm total}$ and ${\cal F}_{\rm H_2}$ for models 
A and C and models B and G clearly shows that the momentum transfer efficiency 
from the jet to the ambient medium {\em decreases} with increasing density.  
This result is particularly marked in the case of post-shock ambient molecules. 
Although further simulations should be performed to confirm this finding, it 
is physically plausible. Cooling causes the bow shock to be narrower 
(i.e.\ more aerodynamic) than in the adiabatic case, thus reducing its 
cross-sectional area.  Obviously this leads to a reduction in rate at which 
momentum is transferred from the jet to its surroundings. The fact that the 
effect is more marked for post-shock ambient molecules must reflect changes 
in the shape of the bow (as opposed to pure changes in its cross sectional 
area) with increased cooling.  

Since we are simulating systems here which are probably of low density 
in comparison to typical YSO jets, our results suggest that radiative bow 
shocks, from at least heavy and equal density jets (with respect to the 
environment), are not very good at 
accelerating ambient molecules without causing dissociation. This result also
points to the fact that in the case of such jets, the jet may carry much more 
momentum than one might naively estimate based on a rough balance with the 
momentum in any associated observed molecular flow.  

\begin{figure*}
\resizebox{\hsize}{!}{\includegraphics{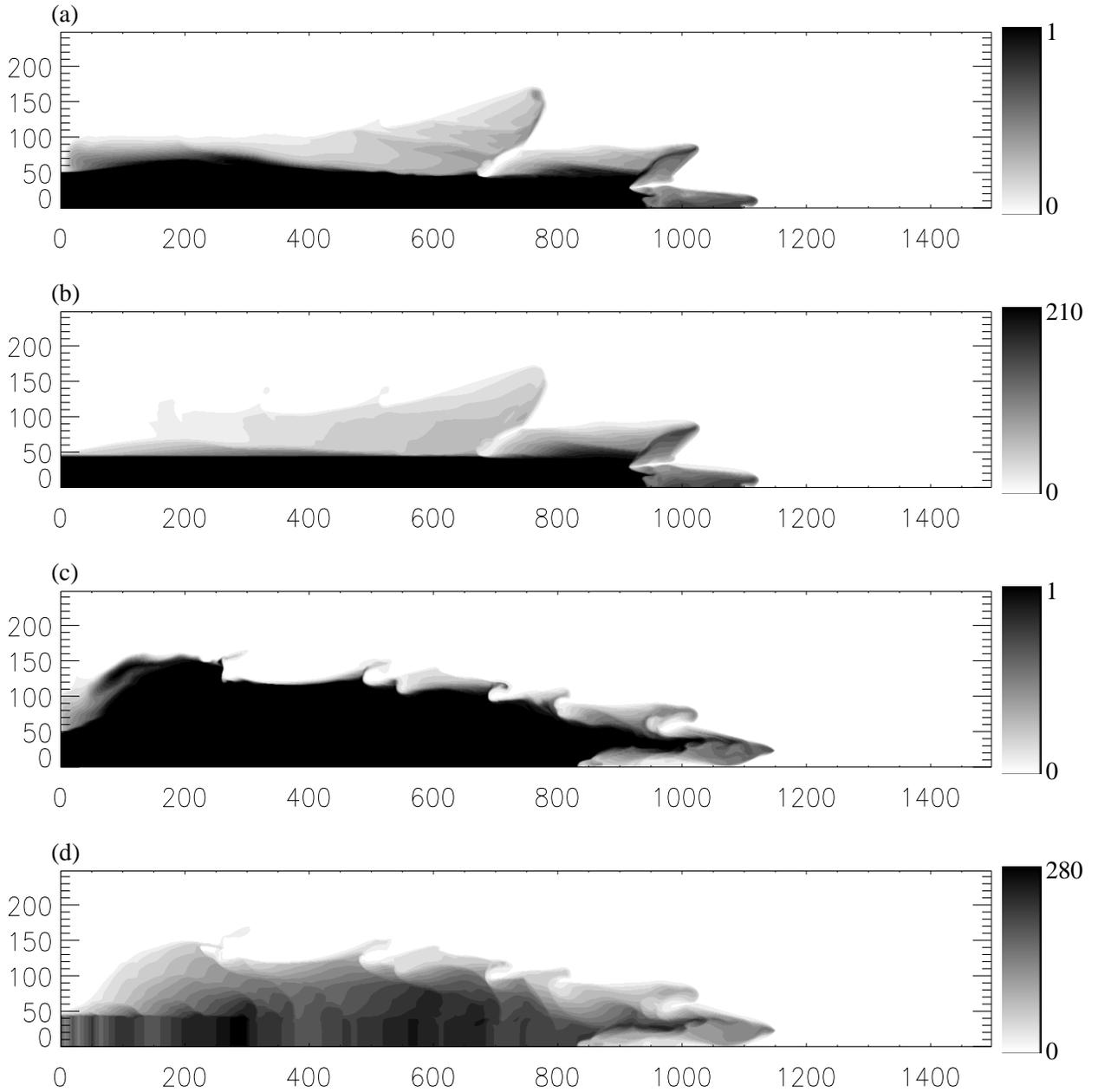}}
\caption{Plots (a) and (b) show the distribution of the jet tracer
variable and $|\vec{u}|$ for simulation C respectively.  Plots (c) and 
(d) show the same quantities for simulation G.  Plots (a) and (c) use a
linear scale and a value of 1 indicates pure jet gas.  Plots (b)
and (d) also use a linear scale and are in units of km s$^{-1}$}
\label{tau_mol}
\end{figure*}

We now turn to differences in the efficiency of momentum transfer in pulsed 
versus steady jets.  The topic of differences in entrainment rates will be 
discussed more fully in \S 3.5. Fig.\ \ref{tau_mol} shows grey-scale plots of 
the distribution of $|\vec{u}|$ and of jet gas for models 
C ($\frac{v_1}{v_0}=0$) and G ($\frac{v_1}{v_0}=0.6$).  Comparison between the
steady jet and the varying velocity jet suggests that momentum is indeed
being forced out of the beam of the varying jet by the internal working
surfaces as predicted for example by Raga et al.\ (\cite{Retal93}).  
It is also interesting to note the similarity between the 
distribution of velocity and the distribution of jet gas in both
simulations.  Moreover it is clear that the momentum leaving the jet beam is 
dumped into jet gas which has been processed through the jet-shock 
and internal working surfaces and now forms a cocoon around the jet itself.
Since this gas is largely atomic (most of it having passed through 
strong shocks), this effect does not directly lead to extra acceleration 
of molecular gas.  However, the ejected momentum could conceiveably pass through
the cocoon of jet gas eventually and go on to accelerate ambient molecules.
The wings of the shocks caused by the internal working surfaces in
the jet have encountered the edge of the cocoon by the end of these 
simulations but, even so, the fraction of momentum in ambient molecules 
varies by not more than 3\% as a result of the velocity variations.
This result was noted by Downes (\cite{downes}) for slab symmetric jets, 
but here we extend this result to cylindrical jets with a variety of 
strengths of velocity variations.  It is also interesting to note that,
from comparisons between simulations G and I, the efficiency of momentum
transfer (in particular to molecular gas) is not very sensitive to 
$\eta$.

\begin{table}[t]
\begin{tabular}{cccc} \hline
Model & ${\cal F}_{\rm H_2}$  & ${\cal F}_{\rm total}$ & $\gamma$ \\ \hline
A & 0.21 & 0.58 & 2.42 \\
B & 0.20 & 0.55 & 2.93 \\
C & 0.08 & 0.36 & 2.37 \\
D & 0.07 & 0.36 & 2.08 \\
E & 0.07 & 0.36 & 1.81 \\
F & 0.09 & 0.33 & 2.31 \\
G & 0.10 & 0.38 & 2.98 \\
G1 & 0.10 & 0.39 & 2.02 \\
G2 & 0.10 & 0.39 & 3.75 \\
H & 0.06 & 0.16 & 1.58 \\
I & 0.08 & 0.21 & 2.44 \
\end{tabular}
\caption{\label{fractions}  The proportion of momentum on the grid
residing in ambient molecules (${\cal F}_{\rm H_2}$) and in all ambient
gas (${\cal F}_{\rm total}$) at $t=300$ yrs.  Also given is the
value of $\gamma$, the power-law index for the mass-velocity relationship}
\end{table}

\subsection{The mass-velocity relationship}

\begin{figure}
\resizebox{\hsize}{!}{\includegraphics{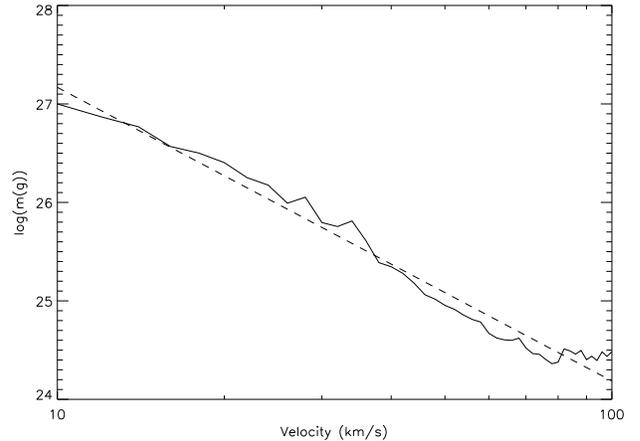}}
\caption{Plot of the relationship between the molecular mass and velocity 
for simulation G at $t=300$ yrs assuming the jet moves at an angle 
of $60^{\circ}$ to the plane of the sky.  Note how a power-law 
(dashed line) fits the data quite well}
\label{m-v}
\end{figure}

We do find a power-law relationship between mass of molecular gas and 
velocity.  If we write
\begin{equation}
m(v) \propto v^{-\gamma}
\end{equation}
then we find that $\gamma$ lies between 1.58 and 3.75 and that $\gamma$ tends
to increase with time, in agreement with Smith et al.\ (\cite{smith}).  
Table \ref{fractions} shows the values of $\gamma$ at $t=300$ yrs for all 
the models.  These values are consistent with observations (e.g.\ 
Davis et al.\ \cite{gam_obs}), and also with the analytical model 
presented in the appendix of Smith et al. (\cite{smith}) for the 
variations of mass with velocity.  However, it is important to 
emphasise that what is actually observed is a variation in CO line
intensity with velocity.  CO line intensity is directly proportional to
mass, in the relevant velocity channel, providing we are in the
optically thin regime and the temperature of the gas is higher than the
excitation temperature of the line (see, e.g.\, McKee et al.\
\cite{mckee}).  Note that Smith et al.\ (\cite{smith}) incorrectly state
that the channel line brightness scales with $v^2\,dm(v)$.  Fig.\ 
\ref{m-v} shows a sample plot of the molecular mass versus velocity for the jet 
moving at an angle of $60^{\circ}$ to the plane of the sky.  We do not 
see the jet contribution in the velocity range chosen here.

The molecular fraction in the jet has a marked influence on the value 
of $\gamma$ predicted by these models as we can see by comparing the 
results for simulations G and G2.  In fact, $\gamma$ increases with 
decreasing molecular abundance in the jet.  This is due to the reduction
in strength of the high velocity jet component.  

It appears that $\gamma$ does not depend in a systematic way on the
amplitude of the velocity variations.  Note also that the introduction of
a wide shear layer dramatically reduces $\gamma$.  This is due to the fact
that more gas is ejected out of the jet beam (because of the more 
strongly paraboloid shape of the internal working surfaces) and this 
accelerates the cocoon gas, leading to a stronger high velocity 
component.  In addition, a wide shear layer causes the bow shock to be
more blunt.  It can be seen from the analytic model of Smith et al.\
(\cite{smith}) that this also leads to a lower value of $\gamma$.
It is interesting to speculate that lower values of gamma, which may be 
more common in molecular outflows from lower luminosity sources (see 
Davis et al.\ \cite{gam_obs}) could result from such flows having a 
higher molecular fraction in their jets and perhaps a wide shear layer.

The behaviour of $\gamma$ with viewing angle is the same as that noted
by Smith et al. (\cite{smith}).  The actual values of $\gamma$ obtained
by these authors are somewhat lower than those obtained here.  However, 
since our initial conditions are so different, and since $\gamma$ is 
dependent on the shape of the bow shock, this discrepancy is not disturbing.

\subsection{H$_2$ proper motions and emissions}

We measured the apparent motion of the emission from the internal
working surfaces.  Near the axis of the jet this emission moves with the
average jet speed (i.e.\ $v_0$), as would be expected from momentum 
balance arguments.  However, there are knots of emission arising from 
the bow shock itself and these move much more slowly ($\sim$5--15\% of 
the average jet speed) with the faster moving knots being closer to the 
apex of the bow.  This is in agreement with the observations of 
Micono et al.\ (\cite{micono}).

\begin{figure*}
\resizebox{\hsize}{!}{\includegraphics{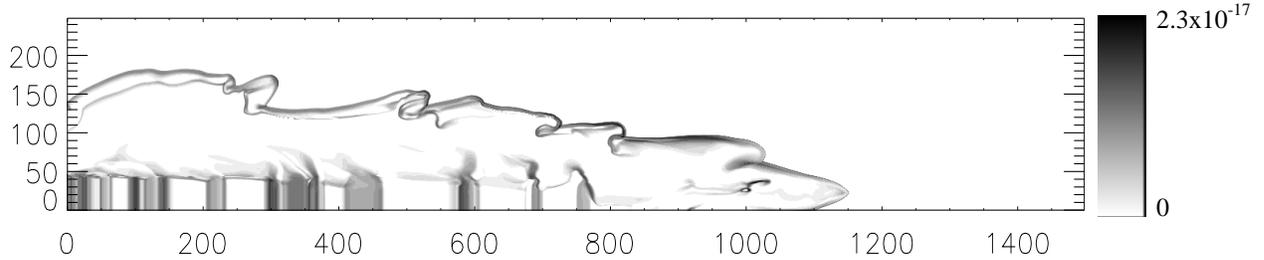}}
\caption{Log-scale plot of the distribution of emission from the 
S(1)1--0 2.12$\mu$ line of H$_2$ for simulation G after 300 yrs.  The scale is 
in units of erg cm$^{-3}$ s$^{-1}$}
\label{h2_em}
\end{figure*}

Fig.\ \ref{h2_em} shows the emission from the S(1)1--0 line of H$_2$
for model G.  There is little emission from the cocoon since the cocoon
gas has been strongly shocked in the jet shock and so is mostly
atomic.  We can also see that the emission becomes more intense as we
move away from the apex of the bow shock, as reported by many authors
(e.g.\ Eisl\"offel et al.\ \cite{eisloffel}).  It is also clear that 
the internal working surfaces in the jet are giving rise to emission in 
this line.  We can see that the emission begins to die away as we move 
away from the jet source.  This is in agreement with observations of, 
for example, HH 46/47 (Eisl\"offel et al.\ \cite{eisloffel}) where 
the emission from the knots appears close to the jet source and then 
fades away.  

This decrease in emission happens for two reasons.  The first is that
the shocks in the jet become weaker as they move away from the source 
simply because the velocity variations, which give rise to the shocks in 
the first place, are smoothed out by the shocks (see, 
for example Whitham (\cite{Whitham}). In addition, the mass flux through 
an individual shock decreases with time due to the divergent nature of the 
flow ahead of each internal working surface.  This means that the emission will
decrease because there is less gas being heated by the
shock.

\subsection{The `Hubble law'}

\begin{figure}
\resizebox{\hsize}{!}{\includegraphics{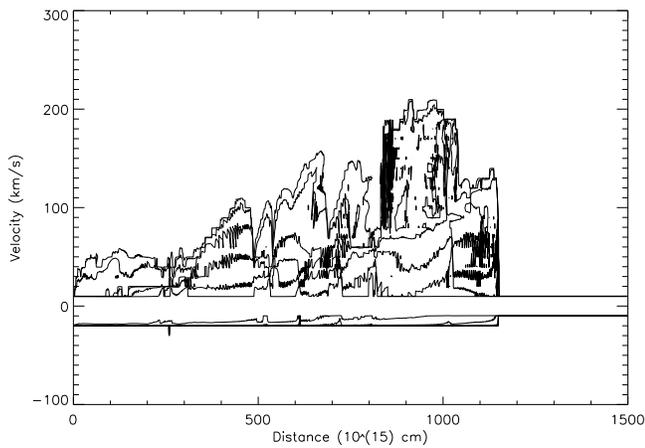}}
\caption{Contour plot of the position-velocity diagram (of mass rather
than emission) for simulation G after 300 yrs assuming the jet moves at an
angle of $60^{\circ}$ to the plane of the sky.  Note the gradual rise of 
the maximum velocity as we move away from the source.  The contours are 
logarithmic running from $3\times10^{21}$ to $3\times10^{25}$ g.  The
contribution from the jet is removed from this diagram to make the
effect clearer}
\label{hub_law}
\end{figure}

We have found that the so-called `Hubble law' (e.g.\ Lada \& Fich
\cite{LF}) is reproduced in these simulations.  Fig.\ \ref{hub_law} 
shows a position velocity
diagram calculated from simulation G assuming that the jet makes an
angle of $60^{\circ}$ to the plane of the sky.  This diagram is based on 
the mass of H$_2$ rather than intensity of CO emission.  There is a 
gradual, virtually monotonic, rise in the maximum velocity.  It is also 
worth noting that near the apex of the bow shock the rise in the maximum 
velocity present becomes steeper.  These properties are related to the 
shape of the bow shock as gas near the apex of the shock is moving 
away from the jet axis at higher speed than that far from the apex.  

\begin{figure}
\resizebox{\hsize}{!}{\includegraphics{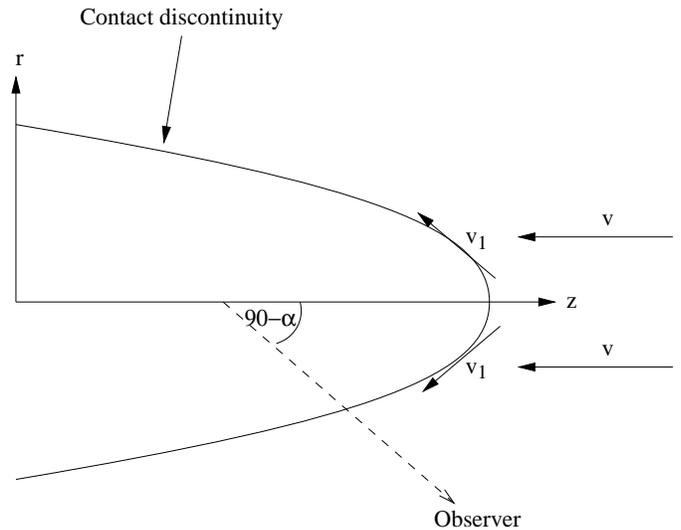}}
\caption{Schematic diagram of the setup used to derive the Hubble law
for position-velocity diagrams.  The fluid motion is shown by vectors.
After contact with the bow shock, the fluid is assumed to flow along the
contact discontinuity with a smaller, but constant, velocity.  See text}
\label{schematic}
\end{figure}

As a very basic model of this, suppose we represent the contact 
discontinuity between the post-shock jet and ambient gas to be an
impermeable body moving with velocity $v$ through a fluid whose 
streamlines will follow the surface of the body.  See Fig.\
\ref{schematic} for a schematic diagram of the system.  Let this surface
be described by the equation
\begin{equation}
z=a-r^s
\end{equation}
\noindent where $s\geq2$ and $a$ is the position of the apex of the bow shock
on the $z$ axis (see, e.g., Smith et al. \cite{smith}).  Since 
the contact discontinuity is a streamline of the flow we get that the
ratio of the $z$-component to the $r$-component of the velocity is simply
\begin{equation}
\label{hub_eqn}
{\cal R}\equiv\frac{-v_z}{v_r}=s\left[a-z\right]^{\frac{s-1}{s}}
\end{equation}
Note that this is the negative of the slope of the bow shock.  This is
because of our choice of the bow shock pointing to the right, and hence
the $z$ component of the velocity will be negative.  If we assume the 
post-shock velocity to be $v_1(z)$ (related to $v$ by the shock
jump conditions), it is simple to show that
\begin{equation}
v_r(z)=\frac{v_1}{\sqrt{1+{\cal R}^2}}
\end{equation}
Finally, after some simple algebra,  we can write down the velocity along 
the line of sight as a function of $z$ by
\begin{equation}
v_{\rm los}(z)=\frac{v_1}{\sqrt{1+{\cal R}^2}}\left\{
	\cos \alpha + \cal{R} \sin \alpha \right\}
\end{equation}
where $\alpha$ is the angle the bow shock makes to the plane of the sky.
In fact we can write down $v_1(z)$ if we assume that the bow shock is a
strong shock everywhere, thus yielding a compression ratio of 4 (from
the shock conditions).  It is easy to derive that 
\begin{equation}
v_1(z) = v \cos\left(\arctan({\cal R})\right) 
	\sqrt{\frac{1}{16}+{\cal R}^2}
\end{equation}

This formula yields a shape for the position-velocity diagram which
suggests the Hubble-law and is similar to diagrams generated 
from these simulations assuming that the flow is in the plane of the sky.
This indicates that the `Hubble-law' effect is, at least partly, an artifact
of the geometry of the bow shock.  

\subsection{Entrainment}

As is clear from Fig.\ \ref{tau_mol} there is not much extra entrainment of
ambient gas resulting from the velocity variations.  If the velocity 
variations were to involve the jet `switching off' for a time comparable 
to the sound crossing time of the cocoon then we would expect ambient 
gas to move toward the jet axis and probably be driven into the 
cocoon when the jet switches on again.  This does not happen in these
simulations where the maximum period of the variations is 50 yrs and the
amplitude is at most 60\% of the jet velocity.  However, there is a
small amount of acceleration of molecular gas at the left-hand
boundary of the grid.  This effect is quite small, but may grow over 
time.  It is also possible, however, that this effect is due simply to 
the reflecting boundary conditions.

These simulations show very little mixing between jet and ambient
gas except very close to the apex of the bow shock.  This means that the
high velocity component of CO outflows often observed along the main
lobe axis (Bachiller \cite{bachiller}) is difficult to explain without 
invoking the presence of CO gas in the jet beam itself just 
after collimation.

\section{Conclusions}

It is now generally agreed that low velocity molecular outflows represent
ambient gas that is somehow accelerated by highly collimated and 
 partly ionized jets, at 
least in the case of low and intermediate mass YSOs. Most of this 
acceleration is thought to be achieved at the head of the jet 
through the leading bow shock (the so-called prompt entrainment mechanism).
In this paper we have examined through many axially symmetric simulations 
the efficiency of the prompt entrainment mechanism as a means of transferring 
momentum to ambient molecular gas without causing dissociation. It is 
found, as one would expect, that the fraction of jet momentum transferred to 
the ambient environment depends on the jet/ambient density ratio. More 
importantly, we see that cooling, which is particularly important at higher 
densities, decreases the fractional jet momentum that goes into 
{\em ambient molecules}. It would seem on the basis of the simulations 
presented here that both heavy and equal density (with respect to the 
environment) jets with radiative cooling have very low efficiencies at 
accelerating ambient molecules without causing dissociation. In part this 
is because cooled jets have more aerodynamic bow shocks than the corresponding
adiabatic ones (i.e.\ they present a smaller cross sectional area to the 
ambient medium). The actual shape, however, of the bow shock also seems to 
to be important as the decrease in momentum transferred to the ambient medium 
seems to affect the acceleration of molecules more so than atoms/ions.  

We have also tested whether pulsed jets are more efficient at transferring 
momentum to the ambient medium than the corresponding steady jet with the same
average velocity. Somewhat surprisingly we found that even relatively large 
velocity variations do not give rise to significant changes in the amount of 
momentum being deposited in ambient gas.  Fundamentally this is because in 
high Mach number jets, even with cooling, there is very little coupling between 
the jet's cocoon and the ``sheath'' (i.e.\ the post-shock ambient gas). This 
lack of coupling is also the reason why turbulent entrainment is not 
significant in YSO jets.  

Our simulations were also used to model the expected variation of mass with 
velocity in molecular flows. We found relatively large $\gamma$ values, 
i.e.\ mass should decline steeply with velocity, in line with observations. 
Interestingly we found that a wide shear layer and an increasing molecular 
component in the jet {\em reduced} $\gamma$. If, as one might expect, such 
conditions are common among outflows from low luminosity YSOs, this could 
explain their observed lower values for $\gamma$. Finally we have shown 
that the so-called Hubble Law for molecular outflows is almost certainly
a local effect in the vicinity of a bow shock.   

\appendix

\section{Overcoming negative pressures}
\label{appendix}

As noted in \S \ref{num_method} it is common for conservative numerical
codes to predict negative pressures under certain conditions.  Flows giving
rise to such problems are usually highly supersonic, and diverging.  The
difficulties occur because the ratio of internal to total energy goes like 
$\frac{1}{M^2}$, where $M$ is the Mach number of the flow.  Therefore,
in a highly supersonic flow, the fractional error required to predict a 
negative pressure is rather small.  The introduction of energy losses 
exacerbates this problem.  This is due to the fact that relatively weakly
diverging flows, for example, can become supersonically diverging when the
system is cooled because the sound speed is reduced.

Schemes which are second order accurate in space tend to produce more 
negative pressures than first order ones.  This is because first order 
schemes dissipate strong features quickly so that strongly diverging flows 
rarely occur.  It seems reasonable, then, to invoke a scheme which is
first order in space whenever a negative pressure is produced as this
will introduce extra dissipation.  This extra dissipation may eliminate the
negative pressure by moving some extra energy from neighbouring cells 
into the problem one.  The scheme used by the code in this work can be 
summarised as follows:
\begin{equation}
\label{scheme}
\vec{U}^{n+1}_i=\vec{U}^n_i-\lambda\left[
^2\vec{F}^{n+\frac{1}{2}}_{i+\frac{1}{2}}-\,^2\vec{F}^{n+\frac{1}{2}}_{i-
\frac{1}{2}}\right]
\end{equation}
where $\vec{U}^n_i$ and $^2\vec{F}^n_i$ are the state vector and second
order flux calculated at $t=n\Delta t$ and $x=i \Delta x$ respectively,
and $\lambda=\frac{\Delta t}{\Delta x}$.  If this scheme produces a
negative pressure at $t=n+1$ then simply apply Eq. \ref{scheme} using
the first order fluxes $^1\vec{F}^{n+\frac{1}{2}}_{i\pm\frac{1}{2}}$ 
instead of the second order ones.  Note that it is necessary to adjust the 
neighbouring cells also so that overall conservation is maintained.

While this fix does not work for all problems, it was found to be
rather effective in the simulations presented here where very strong
rarefactions are produced both at the edge of the jet at the $z=0$
boundary, and within the jet itself where the enforced velocity
variations can cause problems.  This fix can be implemented very easily
by ensuring that the flux vectors around problem cells are stored.  It
has the advantage that it does not involve losing conservation
of any of the physically conserved quantities.  Reducing the scheme to
first order in certain regions of the grid is not a significant problem
because this happens around shocks anyway in order to maintain
monotonicity. 

\begin{acknowledgements} 
We would like to thank C.\ Davis, A.\ Gibb and D.\ Shepard for
interesting discussions on the mass-velocity relationship in molecular
outflows. We would also like to thank Luke Drury for his assistance with
the development of the code and T.\ Downes wishes to acknowledge the support 
of the NWO through the priority program in massively parallel computing. Our 
simulations were carried out on the Beowulf cluster at the Dublin Institute 
for Advanced Studies. Finally we wish to thank the referee, Michael Smith 
for his very helpful comments. 

\end{acknowledgements}

\end{document}